\documentclass[conference]{IEEEtran}
\usepackage{amsmath,amsfonts,amssymb,mathtools}
\usepackage{graphicx}
\usepackage{algorithmic,algorithm,threeparttable}
\usepackage{amsthm,lipsum}
\usepackage[noadjust]{cite}

\newtheorem{prop}{Proposition}

\newcommand{\Fopt}{{\mathbf{F}_\mathrm{opt}}}
\newcommand{\FRF}{{\mathbf{F}_\mathrm{RF}}}

\newcommand{\FBB}{{\mathbf{F}_\mathrm{BB}}}

\newcommand{\WBB}{{\mathbf{W}_\mathrm{BB}}}
\newcommand{\WRF}{{\mathbf{W}_\mathrm{RF}}}
\newcommand{\NRFr}{{N_\mathrm{RF}^\mathrm{r}}}
\newcommand{\NRFt}{{N_\mathrm{RF}^\mathrm{t}}}
\newcommand{\Nt}{{N_\mathrm{t}}}
\newcommand{\Nr}{{N_\mathrm{r}}}

\ifCLASSINFOpdf
\else
\fi

\IEEEoverridecommandlockouts
\IEEEpubid{
	978-1-5090-3009-5/17/\$31.00~\copyright2017 IEEE
}

\begin{document}
\title{Partially-Connected Hybrid Precoding in mm-Wave Systems With Dynamic Phase Shifter Networks}
\author{\IEEEauthorblockN{Xianghao Yu$^*$, Jun Zhang$^*$, and Khaled B. Letaief$^{*\dag}$, \emph{Fellow, IEEE} }
	\IEEEauthorblockA{$^*$Dept. of ECE, The Hong Kong University of Science and Technology, Hong Kong\\
		$^\dag$Hamad Bin Khalifa University, Doha, Qatar\\
		Email: $^*$\{xyuam, eejzhang, eekhaled\}@ust.hk, $^\dag$kletaief@hbku.edu.qa}
	\thanks{This work was supported by the Hong Kong Research Grants Council under Grant No. 16210216. 
	}
}

\maketitle
\begin{abstract}
	Hybrid precoding is a cost-effective approach to support directional transmissions for millimeter wave (mm-wave) communications, and its design challenge mainly lies in the analog component which consists of a network of phase shifters. 
	The partially-connected structure employs a small number of phase shifters and therefore serves as an energy efficient solution for hybrid precoding.
	In this paper, we propose a \emph{double phase shifter} (DPS) implementation for the phase shifter network in the partially-connected structure, 
	which allows more tractable and flexible hybrid precoder design.
	In particular, the hybrid precoder design is identified as an eigenvalue problem. To further enhance the performance, dynamic mapping from radio frequency (RF) chains to antennas is proposed, for which a greedy algorithm and a modified K-means algorithm are developed. Simulation results demonstrate the performance gains of the proposed hybrid precoding algorithms with the DPS implementation over existing ones. Given its low hardware complexity and high spectral efficiency, the proposed structure is a promising candidate for 5G mm-wave systems. 
\end{abstract}


\IEEEpeerreviewmaketitle

\section{Introduction}
The proliferation of smart mobile devices has resulted in an ever-increasing wireless data traffic explosion, which calls for an exponential increase in the capacity of wireless networks \cite{6824752}. Elevating the carrier frequency to millimeter wave (mm-wave) bands has been shown as a promising solution to get rid of the spectrum crunch in wireless systems nowadays \cite{6515173}. 
As a cost-effective way to relieve the prohibitive power consumption of radio frequency (RF) transceiver components at mm-wave frequencies, hybrid precoding is proposed, which adopts only a small number of RF chains to interface a digital precoder at the baseband and an analog RF precoder. 

According to the mapping strategy from RF chains to antennas, hybrid precoding are typically realized by two structures, i.e., fully- and partially-connected structures \cite{7397861}. While the former enjoys the full beamforming gain with each RF chain connected to all the antennas, the latter has drawn much attention recently due to its lower power consumption, lower hardware complexity, and practicability in implementation. In particular, compared to the fully-connected structure, the partially-connected one can reduce the number of phase shifters by a factor of the RF chain number. 
Since the partially-connected structure employs much fewer phase shifters in the analog RF precoder, there will be non-negligible degradation in the analog precoding gain. Hence, how to efficiently use the limited phase shifters in the partially-connected hybrid precoding is still an open problem.

There exist a few studies on hybrid precoding in the partially-connected structure \cite{7006720,6824962,7445130,7397861,han2015large}.
In \cite{7006720,6824962}, codebook-based design of hybrid precoders was presented. While using codebook enjoys a low design complexity, there will be certain performance loss, and how to design the codebook remains to be clarified. By adopting the concept of alternating minimization \cite{7397861} and successive interference cancellation \cite{7445130}, iterative hybrid precoding algorithms were proposed. Nevertheless, one common design difficulty among these approaches is how to tackle the unit modulus constraints induced by the single phase shifter (SPS) implementation, i.e., a single phase shifter is adopted to connect an RF chain with an antenna for which only the phase can be adjusted. Furthermore, these works are either under single-user or narrowband system settings, which is unrealistic since in practice multicarrier transmission is needed to combat the much increased bandwidth, and additional spatial multiplexing gains can be provided by multiuser multiple-input multiple-output (MIMO) techniques.

\IEEEpubidadjcol In \cite{asilomar}, we proposed a novel hybrid precoding implementation, i.e., the double phase shifter (DPS) implementation for the fully-connected structure, which relaxes the unit modulus constraints and leads to a near-optimal performance with effective design methodologies.
In this paper, to reduce the hardware complexity, we extend this implementation to the partially-connected structure for general multiuser OFDM mm-wave systems. While the hardware implementation is similar to our previous work \cite{asilomar}, the design approach is quite different in the partially-connected structure. Specifically, a fixed mapping from RF chains to antennas is firstly considered, where the hybrid precoding problem is identified as an eigenvalue problem with a closed-form solution. Then, to further improve the spectral efficiency, two effective algorithms, i.e., the greedy and modified K-means algorithms, are proposed to dynamically optimize the RF chain-antenna mapping. Simulation results are provided to show the superiority of the proposed approach over existing ones, and the necessity of dynamic mapping in the partially-connected structure.\IEEEpubidadjcol

\section{System Model}
\subsection{Signal Model}
Consider a multiuser OFDM mm-wave MIMO system, where a base station (BS) leverages $\Nt$ transmit antennas to serve $K$ users, each equipped with $\Nr$ receive antennas, and $F$ subcarriers are assumed to be adopted in the downlink transmission. In hybrid precoding, only a small number of RF chains are available, i.e., $N_s\le\NRFt\le\Nt$ and $N_s\le\NRFr\le\Nr$, where $\NRFt$ and $\NRFr$ are the numbers of RF chains at the BS and each user, respectively, and $N_s$ is the number of data streams transmitted to each user on each subcarrier.

The received signal of the $k$-th user on the $f$-th subcarrier is given by
\begin{equation}
\mathbf{y}_{k,f}={\mathbf{W}^H_\mathrm{BB}}_{k,f}{\mathbf{W}_\mathrm{RF}^H}_k\left(\mathbf{H}_{k,f}\sum_{k=1}^K\FRF\FBB_{k,f}\mathbf{s}_{k,f}+\mathbf{n}_{k,f}\right),
\end{equation}
The signal that the BS transmits to the $k$-th user on the $f$-th subcarrier is denoted as $\mathbf{s}_{k,f}$ such that $\mathbb{E}\left[\mathbf{s}_{k,f}\mathbf{s}^H_{k,f}\right]=\frac{1}{KN_sF}\mathbf{I}_{N_s}$. The digital baseband precoder and combiner for each user on each subcarrier are denoted as $\FBB_{k,f}\in\mathbb{C}^{\NRFt\times N_s}$ and $\WBB_{k,f}\in\mathbb{C}^{\NRFr\times N_s}$, respectively. On the other hand, since the transmitted signals for all the users are mixed up by the digital precoder and the analog precoding is a post-IFFT (inverse fast Fourier transform) operation in OFDM systems, the analog RF precoder $\FRF\in\mathbb{C}^{\Nt\times\NRFt}$ is shared by all the users and subcarriers, which differentiates the multiuser OFDM systems from single-user or narrowband systems. Correspondingly, the analog RF combiner for the $k$-th user $\WRF_k\in\mathbb{C}^{\Nr\times \NRFr}$ is shared over all the subcarriers. In addition, $\mathbf{H}_{k,f}$ and $\mathbf{n}_{k,f}$ are the channel matrix and  additive noise vector, respectively. More details of the system model including the mm-wave channel model can be found in \cite[Section II-A]{asilomar}. 

\subsection{DPS Implementation in the Partially-Connected Structure}
\begin{figure}[tbp]
	\centering
	\includegraphics[height=4.5cm]{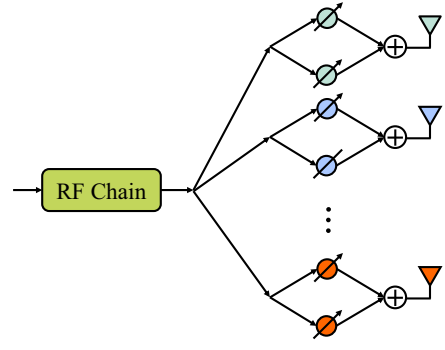}
	\caption{DPS partially-connected structure. Each RF chain is connected to a subset of antennas, and every connection is realized by the sum of two phase shifters.}\label{systemmodel}
\end{figure}

In this paper, we focus on the partially-connected structure for hybrid precoding, given its low hardware complexity and high energy efficiency \cite{7397861}. We extend the DPS implementation of hybrid precoders \cite{asilomar} to the partially-connected structure, as shown in Fig. \ref{systemmodel}, where the beamforming gain from an RF chain to its connected antenna is composed of two phase shifters.

Due to the unit modulus constraint of each phase shifter, the amplitude of the summation involving two phase shifters should be less than two, i.e., the new constraints for the analog RF precoder and combiner are $|\FRF(i,j)|\le 2$ and $|\WRF(i,j)|\le 2$ for all the non-zero entries. This is the key difference compared to the conventional SPS implementation \cite{7397861,6717211}. We shall show that, similar to the fully-connected structure \cite{asilomar}, the DPS implementation also significantly benefits the hybrid precoder design in the partially-connected structure. 

\subsection{Problem Formulation}
It has been shown in \cite{7397861,asilomar,6717211} that minimizing the Euclidean distance between the fully digital precoder and the hybrid precoder is a tractable and accurate approximation for maximizing the spectral efficiency. In this paper, we adopt this alternative objective as our design goal with the following formulation\footnote{In this paper, we focus on the precoder design, and the combiner design problem can be formulated in the same way without the transmit power constraint.}
\begin{equation}\label{problemformulation}
\begin{aligned}
&\underset{\mathbf{F}_\mathrm{RF},\mathbf{F}_\mathrm{BB}}{\mathrm{minimize}} && \left\Vert \mathbf{F}_\mathrm{opt}-\mathbf{F}_\mathrm{RF}\mathbf{F}_\mathrm{BB}\right\Vert _F^2\\
&\mathrm{subject\thinspace to}&&
\begin{cases}
\FRF\in\mathcal{A}\\
\left\|\mathbf{F}_\mathrm{RF}\mathbf{F}_\mathrm{BB}\right\|_F^2\le KN_sF,
\end{cases}
\end{aligned}
\end{equation}
where $\Fopt=\left[\Fopt_{1,1},\cdots,\Fopt_{k,f},\cdots,\Fopt_{K,F}\right]\in\mathbb{C}^{\Nt\times KN_sF}$ is the combined fully digital precoder, and $\FBB=\left[\FBB_{1,1},\cdots,\FBB_{k,f},\cdots,\FBB_{K,F}\right]\in\mathbb{C}^{\NRFt\times KN_sF}$ is the concatenated digital baseband precoder. The second constraint is the transmit power constraint at the BS side. The analog RF precoder $\FRF$ is a common component for all $K$ users and $F$ subcarriers, and is constrained in a candidate set $\mathcal{A}$ determined by the phase shifter implementation and RF chain-antenna mapping, which will be later specified as $\mathcal{A}_\mathrm{f}$ and $\mathcal{A}_\mathrm{d}$ for the fixed and dynamic mapping, respectively.

As problem \eqref{problemformulation} goes, our target is to find an accurate approximation for an arbitrary fully digital precoder. It is also intuitive that the optimal hybrid precoder will lie ``close'' to the fully digital precoder in the Euclidean space. With this formulation, the proposed algorithm can be applied with any fully digital precoder. In this paper, we adopt the classical block diagonal (BD) precoder as the fully digital one, which is asymptotically optimal in the high signal-to-noise ratio (SNR) regime \cite{1261332}.

\section{Hybrid Precoding with Fixed Mapping}\label{IVA}
In \cite{7397861,7445130}, fixed mapping was considered in the partially-connected structure, i.e., each RF chain is connected to a certain number of antennas in a predetermined manner.
The corresponding constraint on the analog RF precoding matrix can be visualized as a set of block diagonal matrices $\mathcal{A}_\mathrm{f}$, where each block is an $\Nt/\NRFt$ dimension vector, i.e.,
\begin{equation}\label{eq22}
\FRF=\mathrm{diag}\left(\mathbf{p}_1,\mathbf{p}_2,\cdots,\mathbf{p}_\NRFt\right),
\end{equation}
where $\mathbf{p}_j=\left[a_{(j-1)\frac{\Nt}{\NRFt}+1},\cdots,a_{j\frac{\Nt}{\NRFt}}\right]^T$. The amplitude of the analog precoding gain for the $i$-th connection from RF chains to antennas is denoted as $|a_i|\le2$. 
This constraint is in fact redundant in hybrid precoding since one can always normalize the unconstrained $\FRF$ with the factor $\gamma=||\FRF||_{\infty}/2$ to satisfy the constraint $\FRF\in\mathcal{A}_\mathrm{f}$, and multiply the digital baseband precoding matrix $\FBB$ by $\gamma$, which will not affect the objective function.
Furthermore, the transmit power constraint $\left\|\mathbf{F}_\mathrm{RF}\mathbf{F}_\mathrm{BB}\right\|_F^2\le KN_sF$ is also automatically satisfied by the optimal solution of the hybrid precoder, which will be shown in the following parts. Thus, the hybrid precoder design problem with fixed mapping can be recast as
\begin{equation}\label{eq23}
\begin{aligned}
&\underset{\mathbf{F}_\mathrm{RF},\FBB}{\mathrm{minimize}} && \left\Vert \mathbf{F}_\mathrm{opt}-\mathbf{F}_\mathrm{RF}\mathbf{F}_\mathrm{BB}\right\Vert _F^2\\
&\mathrm{subject\thinspace to}&&
\FRF\in\mathcal{A}_\mathrm{f}.
\end{aligned}
\end{equation} 
Note that there is only one non-zero element in each row of the analog RF precoding matrix $\FRF$. Due to this special structure, different vectors $\mathbf{p}_j$ will be multiplied by distinct rows of $\FBB$, which decouples problem \eqref{eq23} into $\NRFt$ subproblems in an RF chain-by-RF chain sense. The optimization of the hybrid precoder for the $j$-th RF chain is given by
\begin{equation}
\mathcal{P}_j:\quad\underset{\{a_i\},\mathbf{x}_j}{\mathrm{minimize}}  \sum_{i\in\mathcal{F}_j}
\left\Vert\mathbf{y}_i-a_i\mathbf{x}_j\right\Vert_2^2,\\
\end{equation}
where $\mathcal{F}_j=\left\{i\in\mathbb{Z}\left|{(j-1)\frac{\Nt}{\NRFt}+1\le i\le j\frac{\Nt}{\NRFt}}\right.\right\}$, $\mathbf{y}_i=\mathbf{F}_\mathrm{opt}^T(i,:)$, and $\mathbf{x}_j=\mathbf{F}_\mathrm{BB}^T(j,:)$. 
\begin{prop}
	The optimal solution to the subproblem $\mathcal{P}_j$ is given by the following closed-form expression.
	\begin{equation}
	\mathbf{x}_j^\star=\boldsymbol{\lambda}_1\left(\sum_{i\in\mathcal{F}_j}\mathbf{y}_i\mathbf{y}_i^H\right),\quad a_i^\star=\frac{\mathbf{x}_j^H\mathbf{y}_i}{||\mathbf{x}_j||_2^2},
	\end{equation}
	where $\boldsymbol{\lambda}_1(\cdot)$ denotes the eigenvector corresponding to the largest eigenvalue of a matrix.
	\begin{IEEEproof}
		We check the first order optimality conditions for $a_i$ as
		\begin{equation}
		\frac{\partial}{\partial a_i}f(a_i,\mathbf{x}_j)=0\Rightarrow a_i^\star=\frac{\mathbf{x}_j^H\mathbf{y}_i}{||\mathbf{x}_j||_2^2}\label{eq3}
		\end{equation}
		where $f(a_i,\mathbf{x}_j)$ is the objective function in subproblem $\mathcal{P}_j$.
		By substituting \eqref{eq3}  into the objective function in $\mathcal{P}_j$, it can be rewritten as
		\begin{align}\label{eq8}
		f(\mathbf{x}_j)&=\sum_{i\in\mathcal{F}_j}\mathbf{y}_i\mathbf{y}_i^H-2\Re\sum_{i\in\mathcal{F}_j}\left(a_i\mathbf{y}_i^H\mathbf{x}_j\right)+\left\Vert\mathbf{x}_j\right\Vert^2_2\sum_{i\in\mathcal{F}_j}|a_i|^2\nonumber\\
		&=\sum_{i\in\mathcal{F}_j}\mathbf{y}_i\mathbf{y}_i^H-\frac{\mathbf{x}_j^H\left(\sum_{i\in\mathcal{F}_j}\mathbf{y}_i\mathbf{y}_i^H\right)\mathbf{x}_j}{\mathbf{x}_j^H\mathbf{x}_j}.
		\end{align}
		The second term in \eqref{eq8} is the Rayleigh quotient,
		which leads to an eigenvalue problem, and therefore the optimal values of $\mathbf{x}_j$ and the objective function in $\mathcal{P}_j$ are given by 
		\begin{equation}
		\mathbf{x}_j^\star=\boldsymbol{\lambda}_1\left(\sum_{i\in\mathcal{F}_j}\mathbf{y}_i\mathbf{y}_i^H\right),\quad
		f^\star=\sum_{i\in\mathcal{F}_j}\mathbf{y}_i\mathbf{y}_i^H-\lambda_1,
		\end{equation}
		which completes the proof.
	\end{IEEEproof}
\end{prop}
 From equations \eqref{eq3} and \eqref{eq8}, we can obtain that
\begin{equation}\label{eq29}
\begin{split}
&\left\Vert\Fopt-\FRF\FBB\right\Vert_F^2=\left\Vert\Fopt\right\Vert_F^2-\sum_{j=1}^\NRFt\lambda_j\\
=&\left\Vert\Fopt\right\Vert_F^2-\sum_{j=1}^\NRFt\left(||\mathbf{x}_j||_2^2\sum_{i\in\mathcal{F}_j}|a_i|^2\right)\\
=&\left\Vert\Fopt\right\Vert_F^2-\left\Vert\FRF\FBB\right\Vert_F^2\ge0\\
\Rightarrow&\left\Vert\FRF\FBB\right\Vert_F^2\le\left\Vert\Fopt\right\Vert_F^2\le KN_sF,
\end{split}
\end{equation}
which means that the transmit power constraint is naturally satisfied by the optimal solutions.
While we fixed the mapping strategy as shown in \eqref{eq22}, the proposed design approach is applicable to an arbitrary mapping strategy.

To thoroughly eliminate the interuser interference, similar to \cite{asilomar}, we cascade an additional BD precoder at the baseband based on the effective channel including the hybrid precoder and physical channel. The same approach will also be adopted for the dynamic mapping strategy in the next section.

\section{Hybrid Precoding with Dynamic Mapping}
In contrast to the fully-connected structure that fully utilizes all the connections from RF chains to antennas, the partially-connected structure will induce non-negligible performance loss \cite{7397861}. In this section, we propose to improve the performance of the partially-connected structure by optimizing the mapping strategy, i.e., we will dynamically determine for each RF chain which antennas it should be connected. The dynamic mapping problem can be written as 
\begin{equation}
\begin{aligned}
&\underset{\mathbf{F}_\mathrm{RF},\FBB}{\mathrm{minimize}} && \left\Vert \mathbf{F}_\mathrm{opt}-\mathbf{F}_\mathrm{RF}\mathbf{F}_\mathrm{BB}\right\Vert _F^2\\
&\mathrm{subject\thinspace to}&&
\FRF\in\mathcal{A}_\mathrm{d},
\end{aligned}
\end{equation} 
where $\mathcal{A}_\mathrm{d}$ is a set of matrices for which every row only has one non-zero entry, i.e., $\left\Vert\FRF(i,:)\right\Vert_0=1$, meaning that each antenna can only be connected to one RF chain.
As indicated by equation \eqref{eq29}, once the mapping is fixed, the optimal value of the objective function in \eqref{eq23} is 
\begin{equation}
\left\Vert\Fopt\right\Vert_F^2-\sum_{j=1}^\NRFt\lambda_1\left(\sum_{i\in\mathcal{D}_j}\mathbf{y}_i\mathbf{y}_i^H\right).
\end{equation}
Hence, when we have the freedom to design the mapping strategy from RF chains to antennas, the design target is to seek the mapping that maximizes the sum of the largest eigenvalues, given by
\begin{equation}\label{eq31}
\begin{aligned}
&\underset{\{\mathcal{D}_j\}_{j=1}^\NRFt}{\mathrm{maximize}} && \sum_{j=1}^\NRFt\lambda_1\left(\sum_{i\in\mathcal{D}_j}\mathbf{y}_i\mathbf{y}_i^H\right)\\
&\mathrm{subject\thinspace to}&&
\begin{cases}
\cup_{j=1}^\NRFt\mathcal{D}_j=\left\{1,\cdots,\Nt\right\}\\
\mathcal{D}_j\cap\mathcal{D}_k=\emptyset,\quad\forall j\ne k,
\end{cases}
\end{aligned}
\end{equation} 
where $\mathcal{D}_j$ is the dynamic mapping set containing the antenna indices that are mapped to the $j$-th RF chain. The dynamic mapping problem is a combinatorial problem and the optimal solution can be given by exhaustive search with an extremely huge number of possible mapping strategies as $\frac{1}{(\NRFt)!}\sum_{k=0}^\NRFt(-1)^{\NRFt-k}\binom{\NRFt}{k}k^\Nt$, which prevents its practical implementation. Therefore, we first propose a greedy algorithm to solve the problem.

In each iteration of the greedy algorithm, we add one connection between an antenna and an RF chain, which maximizes the increment of the largest eigenvalue. Due to space limitation, the pseudo-code of the greedy algorithm is omitted.
Note that the computational complexity of the algorithm is dominated by the calculation of the largest eigenvalue. In the greedy algorithm, the number of times we need to perform the eigenvalue decomposition (EVD) is $\mathcal{O}\left(\Nt\NRFt(1+\Nt)/2\right)$, which is a quite large number especially when large-scale antenna arrays are leveraged in mm-wave MIMO systems. To relieve us from the high computational complexity, we then propose a modified K-means algorithm to solve the dynamic design problem \eqref{eq31}.

We reconsider problem \eqref{eq31} as follows. The problem is equivalent to classifying $\Nt$ vectors (antennas) into $\NRFt$ clusters (RF chains). K-means, aiming at partitioning the observation vectors into $K_\mathrm{cl}$ clusters, is a prevalent approach for cluster analysis in data mining, where $K_\mathrm{cl}$ is a predefined parameter, and turns out to be suitable for problem \eqref{eq31}. 
In the classical K-means algorithm, the objective is to minimize the distortion function which is the sum of the Euclidean distances from each observation vector to the cluster centroid it belongs to. 
However, this distortion function cannot be directly adopted to solve the dynamic mapping design problem \eqref{eq31} since the objectives are quite different. In \eqref{eq31}, the objective is to maximize the sum of the largest eigenvalues of the covariance matrices of each cluster. Therefore, we propose to modify the distortion function in the K-means algorithm as
\begin{equation}\label{eq33}
D^\prime(\mathbf{y}_i,\mathbf{x}_j)=\sum_{j=1}^\NRFt\frac{\mathbf{x}_j^H\left(\sum_{i\in\mathcal{D}_j}\mathbf{y}_i\mathbf{y}_i^H\right)\mathbf{x}_j}{\mathbf{x}_j\mathbf{x}_j^H}.
\end{equation}
The modified distortion function is the sum of Rayleigh quotients of the covariance matrices of each cluster, whose optimal value is the sum of the largest eigenvalues when we maximize \eqref{eq33} over $\mathbf{x}_j$. The overall clustering optimization problem can be written as
\begin{equation}\label{eq34}
\begin{aligned}
&\underset{\{\mathcal{D}_j,\mathbf{x}_j\}_{j=1}^\NRFt}{\mathrm{maximize}} && \sum_{j=1}^\NRFt\frac{\mathbf{x}_j^H\left(\sum_{i\in\mathcal{D}_j}\mathbf{y}_i\mathbf{y}_i^H\right)\mathbf{x}_j}{\mathbf{x}_j\mathbf{x}_j^H}\\
&\mathrm{subject\thinspace to}&&
\begin{cases}
\cup_{j=1}^\NRFt\mathcal{D}_j=\left\{1,\cdots,\Nt\right\}\\
\mathcal{D}_j\cap\mathcal{D}_k=\emptyset,\quad\forall j\ne k,
\end{cases}
\end{aligned}
\end{equation} 
We propose to adopt alternating maximization to solve this problem, which alternately updates the clustering and centroids when the other one is fixed. This approach results in closed-form solutions for the two update procedures.

In the clustering update, we allocate each vector to the cluster whose centroid has the largest inner product with it, i.e., allocate $\mathbf{y}_i$ to the $j^\star$-th cluster, where
\begin{equation}\label{clustering}
j^\star =\arg\underset{j}{\max}\quad\left|\mathbf{y}_i^H\mathbf{x}_j\right|^2.
\end{equation}
In the centroid update, the optimization of the centroids is equivalent to maximizing the Rayleigh quotients for each cluster, whose optimal solution is simply given by the eigenvector corresponding to the largest eigenvalue, i.e.,
\begin{equation}\label{centroid}
\mathbf{x}_j^\star=\boldsymbol{\lambda}_1\left(\sum_{i\in\mathcal{D}_j}\mathbf{y}_i\mathbf{y}_i^H\right).
\end{equation}
Now we have the modified K-means algorithm, which is summarized as \textbf{Algorithm 1}.
\begin{algorithm}[t]
	\caption{Modified K-means Algorithm for Dynamic Mapping}
	\begin{algorithmic}[1]
		\REQUIRE
		$\left\{\mathbf{y}_i=\mathbf{F}_\mathrm{opt}^T(i,:)\right\}_{i=1}^\Nt$
		\STATE Construct the initial centroids $\{\mathbf{x}_j\}_{j=1}^\NRFt$;
		\REPEAT
		\STATE Fix the centroids, allocate $\{\mathbf{y}_i\}_{i=1}^\Nt$ to the clusters according to \eqref{clustering};
		\STATE Optimize the centroids $\{\mathbf{x}_j\}_{j=1}^\NRFt$ using \eqref{centroid} when the clustering is fixed;
		\UNTIL convergence;
		\STATE Calculate $\mathbf{F}_\mathrm{BB}^T(j,:)=\mathbf{x}_j=\boldsymbol{\lambda}_1\left(\sum_{i\in\mathcal{D}_j}\mathbf{y}_i\mathbf{y}_i^H\right)$ for $j\in\{1,\cdots,\NRFt\}$;
		\STATE Compute the additional BD precoder at the baseband to cancel the interuser interference \cite{asilomar}.
		\STATE  For the digital precoder at the transmit end, normalize
		$\widehat{\mathbf{F}}_\mathrm{BB}=\frac{\sqrt{KN_sF}}{\left\Vert\mathbf{F}_\mathrm{RF}\mathbf{F}_\mathrm{BB}\right\Vert_F}\mathbf{F}_\mathrm{BB}$.
	\end{algorithmic}
\end{algorithm}

Note that Steps 3 and 4 both give the globally optimal solutions to the clustering and centroid. Hence, the algorithm will converge to a stationary point since it is a two-block coordinate descent procedure \cite{grippo2000convergence}. 

Recall that EVD is the dominant part of the computational complexity in dynamic mapping design. In each alternating iteration in the modified K-means algorithm, $\NRFt$ times of EVD are needed and therefore the overall times are $\mathcal{O}(N\NRFt)$, where $N$ is the iteration number. For practical settings in Section \ref{SecV}, the modified K-means algorithm will typically converge within 10 iterations, which is much less than $\Nt(1+\Nt)/2$ and thus results in significant complexity reduction compared to the greedy algorithm.

\section{Simulation Results}\label{SecV}
In this section, we numerically evaluate the performance of the proposed design approaches for hybrid precoding in multiuser OFDM mm-wave MIMO systems, where the DPS implementation is adopted and $F=128$ subcarriers are assumed.
All simulation results are averaged over 1000 channel realizations.

\begin{figure}[tbp]
	\centering
	\includegraphics[height=5.5cm]{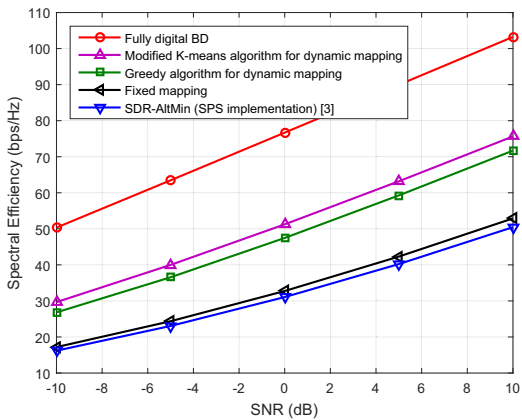}
	\caption{Spectral efficiency achieved by different hybrid precoding algorithms in the partially-connected structure when $\Nt=256$, $\Nr=16$, $K=4$, and $N_s=2$.}\label{fig4}
\end{figure}

Fig. \ref{fig4} shows the performance of the proposed design approaches with the minimum numbers of RF chains, i.e., $\NRFt=KN_s$ and $\NRFr=N_s$.
We see that, due to the sharply reduced number of phase shifters, the partially-connected structure does entail non-negligible performance loss compared to the fully digital one. Furthermore, although the computational complexity of the proposed algorithms is significantly reduced compared to the SDR-AltMin algorithm with the SPS implementation in \cite{7397861}, simply doubling the number of phase shifters with a fixed mapping only provides little performance gain over the conventional SPS implementation. However, Fig. \ref{fig4} demonstrates that dynamic mapping is able to shrink the gap between the fixed mapping and the fully digital precoding by half.

We also compare the performance of the proposed greedy and modified K-means algorithms for the dynamic mapping in the partially-connected structure. Fig. \ref{fig4} illustrates that
the modified K-means algorithm steps up as an excellent low-complexity dynamic mapping algorithm.

\begin{figure}[tbp]
	\centering
	\includegraphics[height=5.5cm]{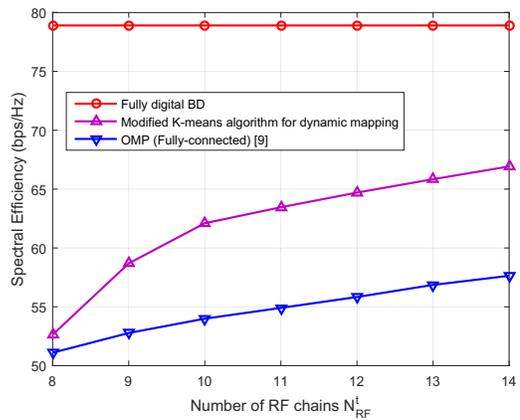}
	\caption{Spectral efficiency achieved by different precoding algorithms for different numbers of RF chains $\NRFt$, given $\Nt=144$, $\Nr=16$, $\NRFr=N_s$, $K=4$, $N_s=2$, and $\mathrm{SNR}=5$ dB.}\label{fig2}
\end{figure}

Fig. \ref{fig2} demonstrates the effect of increasing RF chains in different hybrid precoding structures. Compared with the fully-connected structure where orthogonal matching pursuit (OMP) is adopted as a prevalent low-complexity hybrid precoding algorithm \cite{6717211}, the proposed DPS implementation provides a higher spectral efficiency even with the partially-connected structure, and the performance increases faster as the number of RF chains goes larger. This phenomenon has not been revealed in existing works \cite{7397861,6717211}, which indicates that the DPS implementation with a reasonable number of RF chains is the preference for the partially-connected structure.

\section{Conclusions}
This paper demonstrated the advantages of the recently proposed DPS implementation of hybrid precoding in the partially-connected structure. It showed that, with a delicate design of the mapping strategy and a reasonable number of RF chains, the DPS partially-connected structure can achieve satisfactory performance. Considering its low power consumption and hardware complexity, this structure has the potential of serving as an economic and practical hybrid precoding solution for 5G mm-wave systems.

\bibliographystyle{IEEEtran}
\bibliography{bare_conf}
\end{document}